\documentclass[aps,preprint]{revtex4} 
\usepackage [dvips]{graphicx}

\newcommand{\be}{\begin{equation}}
\newcommand{\ee}{\end{equation}}
\newcommand{\ba}{\begin{eqnarray}}
\newcommand{\ea}{\end{eqnarray}}
\newcommand{\bo}{$^{10}$B }
\newcommand{\nit}{$^{14}$N }
\newcommand{\ca}{$^{14}$C }
\newcommand{\caa}{$^{12}$C }
\newcommand{\bi}{\bibitem}
\newcommand{\pr}[1]{Phys. Rev. Lett. {\bf#1}}
\newcommand{\pc}[1]{Phys. Rev. {\bf#1}}
\newcommand{\np}[1]{Nucl. Phys. {\bf#1}}
\begin{document}
\title{T=0 effective interaction in \nit  and \bo.}
\author{N. Vinh Mau}
\affiliation{Institut de Physique Nucl\'eaire, IN2P3-CNRS,
 Universite Paris-Sud, F-91406 Orsay Cedex, France}

\begin{abstract}
We have calculated the 1$^+$ and 3$^+$, T=0 states in \nit and \bo. In a
neutron-proton RPA model   these two nuclei are described by the same set
of equations. We first show that a bare Minnesota
interaction leads to too weak binding in both nuclei. Furthermore it does
not produce a 
 3$^+$  ground  state  in  \bo as it should. Including medium 
  effects as an exchange of phonons between
the neutron-proton pair  cures the desagreement in \nit but still gives a
1$^+$ ground state in \bo with the 3$^+$ as an excited state. The same study
with a Gogny effective interaction reproduces nicely the properties of both
nuclei: same agreement in \nit  as previously when medium effeccts were
introduced but now the 3$^+$ in \bo becomes the ground state. This
success suggests that through its density dependent term the Gogny
interaction takes account of the presence of a three-body force which, in a
shell model calculation, has
been shown to be essential to give a 3$^+$ ground state in \bo.   
\end{abstract}

\maketitle
\section{Introduction.}

The T=0 neutron-proton pairing correlations in N=Z nuclei have
motivated a number of theoretical and experimental works. In odd-odd N=Z 
nuclei,
T=0 and T=1 states are both present so that they offer a good place to
make a comparative study of the two types of correlations. We have focused
our attention on the two odd-odd nuclei of \bo (N=Z=5)  and \nit  (N=Z=7) 
for several reasons. First  \nit and
\bo  have a 1$^+$, T=0 and a 3$^+$ T=0 ground states
respectively
while the 0$^+$, T=1 states are excited states  and therefore 
suggesting stronger correlations in the T=0 channel.
Furthermore because of charge invariance of the nucleon-nucleon interaction,
the neutron-proton and neutron-neutron effective interactions should
be the same in the T=1 channel.
Therefore an effective interaction and a nuclear model able to explain the
0$^+$, T=1 states in a core plus two neutrons system (like \ca) should also
describe well the same T=1 states in the core plus a neutron-proton
pair (like \nit). We
have then an opportunity to make a further test of our previous work
in collaboration with
 J.C. Pacheco on N=8 nuclei \cite{pv}.  With the Gogny effective 
interaction \cite{goa,gob}or its zero range density dependent 
substitute \cite{sca},
and in a  two-neutron RPA model we were able to
reproduce the 0$^+$  states of \ca-$^{10}$C, $^{12}$Be-$^{8}$Be
and $^{11}$Li-$^7$Li. The nucleus \nit ($^{10}$B)
described as a core of $^{12}$C plus (minus) a neutron-proton pair is the
analog of \ca ($^{10}$C) described as a core of $^{12}$C plus (minus)
a neutron-neutron pair  
 and our
first aim is to check if the same interactions and the same RPA model where
the two-neutron pair is replaced by a neutron-proton pair, will lead
also to a good description of the  0$^{+}$ states in \nit and \bo. 

 In the two-nucleon RPA
model, successful in the description of the T=1 states, we calculate the T=0,
1$^+$ and 3$^+$ states in \nit and \bo. First we perform  the calculation 
with a simple zero range force fitted to reproduce the deuteron binding energy.
  The results obtained
with this simple  potential show a too weak binding. 
Because a zero range force 
is a crude simplification of a realistic one, we have made the same 
calculation with the Minnesota bare interaction \cite{min} which has one 
short range 
component, two long range components  and all exchange terms. The results
are very close to the previous ones and show again a lack of binding. 
It is not surprising that bare
interactions could not describe well bound many-body systems 
 for which we know that medium effects are important. 
Medium effects on the bare nucleon- nucleon  interaction may arise, at least
partly, from the exchange of phonons between the two-nucleon pair. 
Indeed strong
two-body correlations in nuclei manifest themselves as very collective
vibrationnal states  at low excitation energies. These vibrations, or phonons,
may be exchanged between the two nucleons and induce a modification of the
nuclear interaction. The presence of these phonons, besides their
effect on the nuclear two-body force, has an other manifestation by
modifying the average interaction of a single nucleon with the core due to the
coupling between the nucleon and the phonons of the core.
Such couplings
modify the Hartree-Fock average potential  and  are so strong in
nuclei like $^{11}$Be and $^{10}$Li that they are, at least partly, 
responsible for the
inversion of the 1/2$^+$ and 1/2$^-$ states  \cite{sa,vm,br}. 
Both effects, on the 
two-nucleon force and on the nucleon-core interaction  will be included in our
two-particle RPA model for the description of \nit and \bo.  

The spectra obtained in $^{14}$N and $^{10}$B will be compared to those
derived with the phenomenological T=0 Gogny effective interaction
\cite{goa,gob}. From this comparison and comparison with shell model
calculations \cite{arn} using two-body and three-body interactions, we will
try to understand better what is implicitly contained in this empirical
force. 

In section II we present briefly the two-particle RPA model applied to a
neutron-proton pair and precise
our choice of single neutron and proton basis and our choice of two-body
interactions. In section III we  first present the results for the $0^+$, 
T=1 states. Then we show the results  obtained for the T=0 states with the  
two bare nucleon-nucleon interactions and  discuss the  contribution of 
 phonon exchanges between the neutron-proton pair. 
 Then we make the same study with   the
effective Gogny interaction. Comparing  our results with those
of shell model calculations we can give a qualitative interpretation of the
medium effects contained in this very successful effective interaction.
Section IV is devoted to our general conclusions.

 \section{The neutron-proton RPA model}

We describe \nit and \bo as a core of $^{12}$C in its ground state plus or
minus a neutron-proton pair respectively and define two-body amplitudes as:

\ba
X^+_a&=&\langle ^{14}N|A^+_a|^{12}C\rangle \\
Y^+_\alpha&=&\langle^{14}N|A^+_\alpha|^{12}C\rangle\\
X^-_\alpha&=& \langle^{10}B|A_\alpha|^{12}C\rangle\\
Y^-_a&=&\langle^{10}B |A_a|^{12}C\rangle
\ea
where a,b\ldots  and $\alpha$,$\beta$,\ldots  represent configurations where
 the neutron and the proton are respectively in states unoccupied and
occupied  in the Hartree Fock ground state of the \caa core. A$^+_{a(\alpha)}$
(A$_{a(\alpha)}$) are
operators which create (annihilate) a neutron-proton pair coupled to  given
spin and isospin (for simplicity we omit them in our equations). $|^{12}C>$
represents the correlated ground state of the $^{12}C$-core.  
Note that the Y-amplitudes,named the small RPA amplitudes, would be zero for
an uncorrelated core. 

The two-nucleon RPA model has been described and used in a number of papers
and we  remind briefly that:

-the same system of equations determines the amplitudes and energies of \nit
and \bo which then are not independent. For a given spin and isospin these
equations write as:
\ba
(\Omega -\epsilon_a)x_a-\sum_b<a|{V}|b>x_b-\sum_{\beta}
<a|{V}|\beta>x_\beta=0\\
(\Omega-\epsilon_\alpha)x_\alpha+\sum_b<\alpha|{V}|b>x_b+\sum _\beta 
<\alpha|{V}|\beta>x_\beta=0
\ea
where the eigenvalues,$x_a$ and $x_\alpha$, are related to the amplitudes of
eqs.(1-4) as explained below. The $\epsilon_a$ and $\epsilon_\alpha$ are
the unperturbed energies of the neutron-proton pair in states a and $\alpha$
respectively.The matrix elements of the neutron-proton
interaction V have to be antisymmetrised.  

-if the model subspace contains N configurations a,b,..  and M
configurations $\alpha$, $\beta$ \ldots , the RPA equations
have N+M eigenstates with eigenvalues $\Omega$ and eigenvectors x$_a$ and
x$_{\alpha}$. N of them correspond to \nit with:

\ba
&~~&E_n(^{14}N)-E_0(^{12}C)=\Omega_n \nonumber \\
&~~&X^{+(n)}_a=x_a^{(n)}\\
&~~&Y^{+(n)}_\alpha=x_\alpha^{(n)}\nonumber
\ea
and M to \bo with:
\ba
&~~&E_m(^{10}B)-E_0(^{12}C)=-\Omega_m \nonumber \\
&~~&X^{-(m)}_\alpha=x_\alpha^{(m)}\\
&~~&Y^{-(m)}_a=x_a^{(m)}\nonumber
\ea
E$_0$(\caa) is the ground state energy of the $^{12}$C-core.
 The separation into two sets of solutions is unambigous    and based on
 energy considerations and relative values of $x_a $ and $x_{\alpha}$  
 amplitudes.

 - the amplitudes x$_a$ and x$_\alpha$ are normalised according to:
 
 \ba
 \sum_a |x_a|^2-\sum_\alpha|x_\alpha|^2&=&1 \;\;\; for \;\;^{14}N\\
                                       &=&-1 \;\;\; for \;\;^{10}B
\ea

-the only inputs of the calculation are the individual neutron and proton
energies and the effective two-body interaction.

\subsection{Choice of the single particle basis}

We make a semi-phenomenogical approach to the RPA model. We replace the
Hartree Fock average potential by a Saxon-Woods potential plus a spin-orbit
force plus a phenomenological surface potential  fitted to
reproduce 
the experimental single neutron  energies in the field of \caa 
and write the one
neutron hamiltonian as: 
\be
h_n=t_n+V_0\left(f(r)-0.44 r_0^2({\bf l.s}) \frac{1}{r} \frac{df(r)}{dr}\right
)+\delta V_n
\ee
where:
\be
f(r)=[1+exp(\frac{r-R_0}{a})]^{-1}
\ee
with V$_0$=-50.5 MeV, a=0.75 fm,  R$_0$=r$_0$(12)$^{1/3}$ with r$_0$=1.27 fm.
The last term
$\delta V_n$ is added to simulate medium effects due to the coupling of the
neutron with the phonons of the core.. The
shape of $\delta V_n$ is suggested by a semi-microscopic calculation 
of neutron-
phonon couplings \cite{vm,vp}. This contribution to the average one body potential
depends on the neutron state and is written as:
\be
\delta V_n= \alpha_n\left(\frac{df(r)}{dr}\right)^2
\ee

\begin{table}
\begin{center}
\begin{tabular}{|c|cc|cc|cc|cc|} \hline
 &\multicolumn{2}{c|}{1p$_{3/2}$}&
 \multicolumn{2}{c|}{1p$_{1/2}$}&
\multicolumn{2}{c|}{2s}&
\multicolumn{2}{c|}{1d$_{5/2}$}\\ 
  &exp.&cal.&exp. &cal. &exp. &cal. & exp. & cal.\\ \hline
neutrons& -18.72&-18.71&-4.95 & -4.95 & -1.85 & -1.86 & -1.1 &-1.1\\
protons&-15.96&-15.54& -1.94 & -1.98 & 0.42 & 0.25 & 1.61 & 1.35 \\ \hline
\end{tabular}
\caption{Experimental and calculated  individual energies (in MeV) for
neutrons and protons.}
\end{center}
\end{table}

The coefficients $\alpha_n$ are fitted on the experimental neutron energies
for 1p$_{3/2}$, 1p$_{1/2}$, 2s and 1d$_{5/2}$ states and put to zero for
higher states. The proton energies are calculated by adding the
Coulomb potential to the neutron hamiltonian of eq.(11). In Table I are
given the experimental energies  and
the corresponding calculated energies for the lowest, known,
neutrons and protons states. The results presented below have been obtained
with the 1p$_{3/2}$ proton energy replaced by the experimental value. However
it has no effect on the \nit spectrum and in \bo gives both 1$^+$ and 3$^+$
energies are lowered by about 0.5 MeV  what does not change qualitatively our
discussion of results.
 The effect of
the Coulomb potential on the wave functions is neglected and 
neutron and proton wave functions are assumed to be the same.

\subsection{Choice of two-body interactions.}
 We have first used a zero range density dependent interaction. 
 The general form of such a  force writes as:
\be
V({\bf{r_1,r_2}})=-V_0\left\{1-\eta \left
(\frac{\rho(\bf{(r_1+r_2)}/2)}{\rho_0}\right)^{\alpha}\right \}
\delta(\bf{r_1-r_2})
\ee  

With the parameters of Garrido et al. \cite{scb}, $V_0$=500 MeV.fm$^3$, 
$\alpha$=0.47
,$\eta$=-0.1 we get much too weak binding. These parameters were fitted so
that to reproduce the gap  in nuclear matter calculated using the Paris
potential, then they do not take account of the medium effects which are
expected to be important. Therefore we have proceeded in a
different way. We first use a density independent zero range neutron-proton 
interaction
where the strength $V_0$ is fitted to give the binding energy of deuteron.
The calculation leads to a relation between the T=0 and T=1 strengths given
by \cite{be}: 

\be
V_0(T=0)=V_0(T=1)\left \{ 1-\left (\frac{-\epsilon_b}{\epsilon_c}\right )^{1/2}
Arctg\left (\frac{\epsilon_c}{-\epsilon_b}\right )^{1/2}\right \}^{-1}
\ee
where $\epsilon_b$ is the deuteron binding energy and  $\epsilon_c$ 
the cut-off on
the   nucleon  energies which in a nucleus should be counted relative to
the bottom  of the average one-nucleon potential. In our calculations we take
for a nucleon in the field of  our $^{12}$C-core
a cut-off of 10 MeV which is equivalent to $\epsilon_c$=60 Mev and 
corresponds to:
\ba 
V_0(T=0)&=&1.36V_0(T=1) \\
        &=&682 MeV.fm^3
\ea
A similar ratio between the strengths of the T=0 and T=1 pairing
interactions has been found by Satula and Wyss \cite{we} in their study of
even-even nuclei.  However a
zero range force can be considered only as a simple substitute to a more
realistic one,  and before introducing medium effects in our calculations
we have tested this substitution and made the calculation with the Minnesota
free nucleon-nucleon force \cite{min}. This force has one short range repulsive
components and two long range attractive ones with all exchange terms and
has been fitted on nucleon-nucleon scattering lengths and deuteron binding
energy.
The results with the two forces show very similar results with the same
underestimation of two-body correlations suggesting the necessity to take
account of medium effects on the neutron-proton interaction. Indeed when two
nucleons are
 added to  a core   their mutual interaction will be modified by the
presence of the other nucleons. Strong two-body correlations which
manifest themselves as very
collective low energy vibrationnal states  can induce a
modification of the interaction  through an exchange of these collective
phonons between the pair of nucleons. Such phonon exchange contribution
to the two-neutron pairing force  have been studied by 
Barranco et al. \cite{bb}
and found to be  responsible
for about half of the gap in the isotopes $^A$Ca, $^A$Ti and $^A$Sn. 
In our case the $^{12}$C core has a
 low 2$^+$, T=0 state at 4.4 MeV with a very strong collective transition 
 amplitude
$\beta_2$=0.6 and a less collective 3$^-$, T=0 state at 9.6 MeV with 
$\beta_3$=0.4 \cite{sag}. We can expect that these two states will give
 most of the effect and we include both of them in our calculation. 

\begin{figure}[h]
\hspace*{-1.cm}
\includegraphics*[scale=0.9]{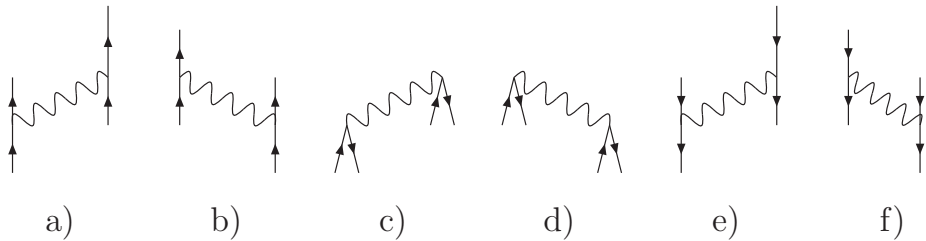}
\caption{Diagrams corresponding to a one phonon exchange between 
neutron-proton pairs appearing in the RPA equations.}
\end{figure}

The diagrams  corresponding to the exchange of phonons 
are represented in Fig.1 for the three types of
matrix elements  entering in the RPA equations. The diagrams a) -b)
concern the matrix elements $<a|V|b>$ while the diagrams c) -d) and e) -f)
concern respectively the matrix elements $<\alpha|V|a>$ and 
$<\alpha|V|\beta>$ of eqs.(5-6)
. The calculation of their
contribution to the RPA matrix elements is given in the Appendix when 
each vertex of the diagrams is
replaced by a phenomenological     anzats. These matrix elements which have
to be added to the bare matrix elements depend on the eigenvalues  of the RPA
equations which therefore are now nonlinear equations and will be solved
by iteration.

 Once one has seen the effect of phonon exchanges on the \nit and \bo
 spectra, the last step is to look if a phenomenological 
 effective interaction such as the Gogny interaction may be interpreted as a
 bare interaction corrected by the diagrams of Fig.1. In the T=0 channel the
Gogny interactions, D1 \cite{goa} and D1S \cite{gob}, have a density 
 dependent component of zero
range. As any zero range interaction, it diverges and its fitted strength
depends on the cut-off of the single nucleon energies. Since our cut-off 
 for neutrons and protons  is quite low (10 MeV) we have used a
slightly increased strength, as will be discussed below.

\section{Results}
 
 Our configuration subspace is restricted to neutron states up to 10 MeV
 and to the corresponding proton states.
 
\subsection{0$^+$, T=1 states}

Previous calculations \cite{pv} have shown that the 0$^+$, T=1 states 
 in the Li, Be
and C isotopes are very well described in the two-neutron RPA model with the
Gogny forces  or their zero range density dependent substitute \cite{sca}.
\nit and \bo are the analogs of $^{14}$C-$^{10}$C where the two-neutron pair
is replaced by a neutron-proton pair added or substracted from a $^{12}$C-
core. Because of charge invariance of strong interactions the same T=1
effective interactions should be able to describe the two kinds of nuclear
systems. The calculation has been performed for \nit and \bo and yields to
the results reported in Table II for the zero range density dependent force
and the Gogny interactions D1 and D1S. We see that the three forces give
 close results as already found in ref[1] for $^{14}$C and a good agreement
 with the measured energies \cite{exp,ex,exp1} even though the excited 
 0$^+$ state is slightly too
 high as it was in $^{14}$C. We may conclude that we have a good test of the
 validity of our two-body RPA model as well as a further confirmation of the
 efficiency of the Gogny effective interactions to describe light nuclei for
 which it was not designed. 

\begin{table}
\begin{center}
\begin{tabular}{|c|cccc|}\hline
 & a)& b)&c)&exp.)\\ \hline
 $^{14}$N &-10.12&-10.18&-10.03&-10.14\\
 & -3.28&-3.10&-3.06&-3.8\\ \hline
$^{10}$B&29.7&29.0&29.24&29.15\\ \hline
\end{tabular}
\caption{Energies(in MeV) of the 0$^+$,T=1 states with respect to the ground
state of $^{12}$C in $^{14}$N and $^{10}$B obtained with: a) the zero range
 density dependent force, b) and c) the D1 and D1S Gogny interaction
respectively. In the last column are given the experimental energies.      }
\end{center}
\end{table}

\subsection{1$^+$ and 3$^+$, T=0 states}

All results of this section are presented in Tables III-V and Figures 2 and
3 for \nit and \bo.

With the zero range interaction of eq.(14) and the parameters of ref.[11],
V$_0$=500 MeV.fm$^3$, $\eta$ =-0.1 and $\alpha$=0.2, our 
results desagree with experimental spectra. In particular the lowest 
1$^+$ state in \nit is above the lowest 0$^+$ T=1 
  which is then the ground state in desagreement with experiments \cite{exp}.

\begin{table}
\begin{center}
\begin{tabular}{|cc|cccccc|} \hline
 & J$^\varpi $&a)&b)& c)&d)&e)&exp. \\ \hline
 $^{14}$N &1$^+$ &-11.7&-11.9&-12.4&-12.8&-12.5\\
 &1$^+$ & -5.8 &-4.61&-5.75 &-6.9&-6.9&-6.3\\
 & 3$^+$ &-3.26&-4.27&-5.55&-7.25&-7.2&-6.05\\ \hline
$^{10}$B&1$^+$ &27.6&28.7&28.1&29.1&28.4&28.1\\
 & 3$^+$ & 30.3 & 29. & 28.6&28.3&27.7  & 27.4 \\ \hline
\end{tabular}
\caption{Energies of the lowest 1$^+$ and 3$^+$,T=0 states in $^{14}$N and
$^{10}$B with respect
to the ground state energy of $^{12}$C obtained with different interactions:
a) the bare zero-range interaction -b) the bare Minnesota interaction -c)the
Minnesota force plus the  exchange of phonons -d) and e) the Gogny force
with respectively the experimental and calculated 1p$_{3/2}$ proton energy  .
In the last column are given the experimental energies.}
\end{center}
\end{table}

 Instead we have first made the calculations using the bare density
 independent zero range neutron-proton interaction with the strength
 $V_0$=682 MeV.fm$^3$ of eq.(17)
 and with the Minesota force \cite{min}. The energies of
 the lowest states referred to the theoretical ground state energy of
 $^{12}$C as defined in eqs.(7-8) are given in the table III.
  We see that the energies are very similar
  for the two bare interactions but that there is a significant desagreement
  with experimental energies for both nuclei. The desagreement is still more
  pronounced for \bo where the ground state is found as a 1$^+$ state while
  experimentally it is a 3$^+$ state. As expected we see clearly
 that a free neutron-proton interaction is not sufficient to give the 
 binding of two  nucleons inside a nucleus. Then
  we have introduced medium effects due to the  exchange of phonons as
 explained in section II.B and in the Appendix and have calculated the induced
 matrix elements
 according to eqs.(21-29).  The results are again very similar for the two
 interactions and we show 
  in the tables and figures those obtained with  the
 Minnesota interaction. The excitation spectra are shown in Fig.2 for
 $^{14}$N and Fig.3 for \bo for the bare and bare plus induced
 interaction.  At the bottom of the figures are   given the
 separation energies of a neutron-proton pair in \nit and $^{12}$C which are
 directly related to the lowest RPA energies in \nit and \bo respectively.
\begin{figure}[h]
\hspace{-2.5cm}
\includegraphics*[angle=-90.,scale=0.5]{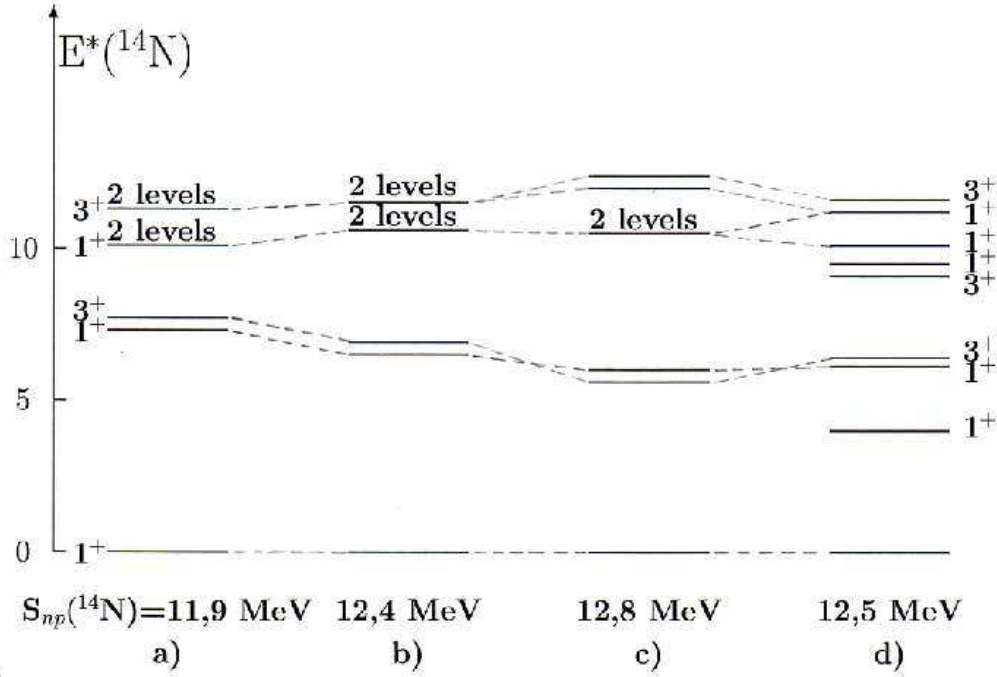}
\caption{Excited spectrum of \nit calculated with: a) the bare Minnesota
interaction-b) the same with the exchange of phonons-c) the Gogny
interaction ( D1S with t$_3$=1600 MeV.fm$^3$. The experimental spectrum is
given in the last column. Below are given the ccalculated and experimental 
neutron-proton separation energies. }
\end{figure}

\begin{figure}[h]
\hspace{-2.5cm}
\includegraphics*[angle=-90.,scale=0.5]{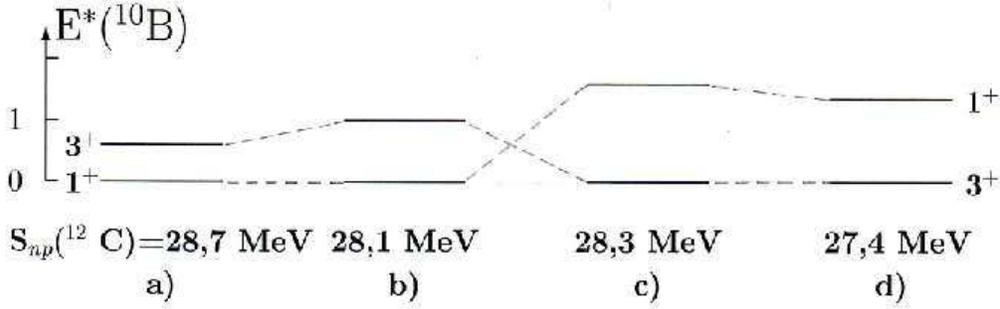}
\caption{Same legend as Fig.2 for \bo. The numbers at the bottom of the
figure are the neutron-proton separation energies, calculated and
experimental, in $^{12}$C.}
\end{figure}

We see an improvement for the \nit excited spectrum  as well as for the
 neutron-proton separation energies.  
 The excitation spectrum of \nit (see  Figure 2)
shows now a quite good agreement with the experimental spectrum.
 Since we take in the definition of our neutron-proton subspace
 nucleon states up to 10 MeV we have in the  higher part of the \nit spectrum
 a large number of
 states but we have given only those  called  pairing vibrational
 states which have amplitudes on several  neutron-proton configurations. 
 The other
 levels have a smaller probability to be seen experimentally \cite{vb}.
For these high energy levels
 we have not calculated the contribution of phonon exchange. Our code is
 inefficient when we have several very close eigenstates but we
expect the induced matrix elements  to be weak because of energy denominators.
 The contribution of phonon exchange is very important for the low
 energy states and   improves significantly the energy
spectrum where now the ground state energy, or equivalently the
neutron-proton separation energy,  is very good. At about  6
MeV  we can reproduce the experimental 1$^+$- 3$^+$ doublet with
the right order of levels
and energies very close to the experimental ones. The main RPA amplitudes
\begin{table}
\begin{center}
\begin{tabular}{|c|c|cccc|c|}\hline
J$^\varpi$ &E$^*$(MeV)&(1p$_{1/2}$)$^2$ & (1d$_{5/2}$)$^2$ & (2s)$^2$ &
(2s,1d$_{5/2}$) & (1p${3/2}$)$^2$\\ \hline
1$^+$ & 0 & 0.88 & 0.24 & 0.1 & & -0.24 \\
1$^+$ & 6.6 & -0.22 & 0.50 & 0.74 & & -0.11\\
3$^+$ & 6.8 & & 0.50 & & 0.78 & -0.13 \\ \hline
\end{tabular}
\caption{RPA amplitudes in \nit }
\end{center}
\end{table}
are given in Table IV for the lowest states . We see that the RPA
Y-amplitudes with two
nucleons in the 1p$_{3/2}$ shell are not negligible what indicates the
presence in the wave function of the $^{12}$C ground state   of
configurations with at least two holes in this shell, in agreement with the
shell model calculation of Cohen and Kurath for example \cite{ck} .    In the
higher part of the spectrum, above 10 MeV, we get a group of two
1$^+$ and two 3$^+$ states which may be identified with the same
experimental group. However we miss the experimental 1$^+$ level at 3.95 MeV
which is very likely  formed mainly of
a $^{12}$C core
excited to its 2$^+$ state at 4.4 MeV coupled to the  n-p pair in
 its 1$^+$  ground state. This is also suggested by  the analysis of
 $^{12}$C ($^6$Li,$\alpha$)$^{14}$N \cite{mk} and $^{16}$O($\gamma$,np)\nit 
 reactions \cite{is,gaa}. 
RPA is not able to describe such state since it relies on the assumption of
an inert core in its ground state. 
In the same way we miss the 1$^+$ and 3$^+$ states at  about 9 MeV
 which again are very likely due to the coupling of the same 2$^+$ in
$^{12}$C with the  neutron-proton pair in  the 1$^+$ or the 3$^+$ excited
states at about 6 MeV. This assumption is supported by the presence 
in the experimental spectrum \cite{exp}of two T=0 levels, a 2$^+$ at 
8.98 MeV and 
a 5$^+$  at 8.96 MeV which very likely belong to the same multiplet.

\begin{table}
\begin{center}
\begin{tabular}{|c|ccc|}\hline
  & (1p$_{1/2}$)$^2_{1^+,0}$&(1p$_{3/2}$)$^2_{1^+,0}$&
  (1p$_{3/2}$)$^2_{3^+,0}$ \\ \hline
  Min& -3.42&-3.74&-5.19\\
  Min'&-3.42&-4.03&-5.49\\
  PWBT & -3.45&-4.16&-6.04\\ \hline
\end{tabular}
\caption{Diagonal matrix elements (in MeV) for the two nucleons in the 
1p$_{1/2}$ or
1p$_{3/2}$ shells coupled to J$^{\varpi}$=1$^+$ or 3$^+$ calculated with the
bare Minnesota interaction (Min), when the phonon exchange contribution is
added (Min') and those fitted in ref.[24] by Warbuton and Brown(PWBT). }
\end{center}
\end{table}

In \bo the results are improved by phonon exchanges but not enough to get
the 3$^+$ state as the ground state. The energy of the 1$^+$ when referred
to the core ground state energy,     is close to the
experimental value      but the 3$^+$ is too high and appears again as an
excited state. This 3$^+$ state  
is a nearly pure (p$_{3/2}$)$^{-2}$ state 
 weakly affected by
RPA correlations because the lowest possible configuration with unoccupied
neutron   and proton unperturbed states (anomalous RPA configuration)
 is the (2s,1d$_{5/2}$) configuration, much higher in energy.
 The correlated energy is therefore mostly due to the diagonal matrix element
$\langle (p_{3/2})^2|V|(p_{3/2})^2 \rangle$ which appears to be too weak as
is discussed now. Indeed we compare our matrix elements with those of 
Warburton and Brown \cite{wa}
 which are determined by least-squares fits to 216 levels in A=10-22
nuclei. They use a (1p,2s,1d) shell model space, compatible with our
subspace, at least for the lowest states.  In Table V are reported  
 their fitted
matrix elements and ours calculated with the Minnesota plus phonon exchange
when the n-p pair is in the 1p-shell. We show the diagonal matrix 
elements for the
n-p pair in the 1p$_{3/2}$ and 1p$_{1/2}$ shells coupled to spin   1$^+$ and
3$^+$ which  are for a large
part responsible for the binding energy of the lowest 1$^+$ states in 
\nit and \bo and the 3$^+$ state in \bo. The same comparison has been made
with the matrix elements fitted by Cohen and Kurath \cite{ck}
in a smaller model space but the conclusion is qualitatively the same. We
 see a  good agreement when the two nucleons are coupled to J=1$^+$ 
 but a too weak binding for J=3$^+$.  
  It shows that something is missing in our force and a possible
reason why it is too weak could be found in more recent shell model
calculations.  Indeed in a no-core shell model and with the 
Argonne V8' two-nucleon force , Aroua et al \cite{ar} get a good description 
of the 1$^+$ ground state of \nit while  Caurier et al.\cite{cau} and Navratil
and Ormand \cite{arn} show that  with the same force and the same model 
it is not possible to get   the
ground state spin of \bo. The authors of ref.[9] show that it is correctly
obtained only when they include
 the Tucson-Melbourne three-body interaction. The necessity to use a
 three-body force in the description of \bo    is also clear from quantum
 Monte Carlo shell model calculations of \bo \cite{pip,wir}  
A genuine three-body force is out of the scoop of the RPA since we work with
two nucleons only. However this three-body term could induce an effective
two-body contribution which is not included when we use the Minnesota force
plus phonon exchange. Intuitively we expect this term to be more important
for two 1p$_{3/2}$-nucleons inside the $^{12}$C, therefore for the
description of \bo,  where they will interact with
a third nucleon off the six other nucleons in the same shell    than for the
two nucleons added in higher shells to describe \nit which will have less
interaction with a third nucleon inside $^{12}$C.  This might explain why our
calculations failed in \bo but are very satisfying in \nit.

At last  we have made the same RPA calculations
with the D1 and D1S Gogny forces \cite{goa,gob}. The two forces give the
same results and we discuss only those obtained with D1S.
 In the T=0 channel this force has a zero range density
dependent component with a strength t$_3$=1390.6 MeV.fm$^3$. Any zero range
interaction is divergent and the fitted strength depends on the cut-off on
single-particle energies. Since our cut-off is somewhat low we have increased
t$_3$ in order to get close to the measured energy of the 1$^+$ 
ground state in \nit.
With  t$_3$=1600 MeV.fm$^3$, a value slightly stronger than the genuine
value, we get a binding energy for the n-p pair of -12.7 MeV instead of
-12.5 MeV, the experimental value.  In  the third columns of Figs.2
and 3 we show the excited spectra
  obtained for \nit and \bo respectively.   We see
 that for \nit the levels are in very close agreement with experiment and
   very close to those obtained with the
 Minnesota interaction plus the phonon exchange contribution with however 
 an inversion of
 the 1$^+$- 3$^+$ states at about 6 MeV. This result suggests that the Gogny
 interaction, often thought as a G-matrix, includes implicitly the
 phonon exchange contribution. In \bo the 3$^+$ state is obtained as the ground
 state as it should, contrarly to what we got previously with an effective
 two-body interaction. 
 The
1$^+$ state is now  the
first excited state with an excitation energy in agreement with the
experimental value and with shell model calculations of ref[9]. Therefore 
 according to our previous discussion on shell model results we may 
 conclude  that,
 through its density dependent term, the Gogny interaction  includes
 implicitly  an effective two-body contribution
coming from three-body forces.However 
the neutron-proton  separation energy in $^{12}$C given as the
difference between the ground state energies in \bo and $^{12}$C is too high
by 0.8 MeV but  a still larger overestimation is observed in the  shell
model calculations (see their tables V and VII). Note that if we use the
calculated energy for a 1p$_{3/2}$ proton (see Table 1) instead of the
experimental energy  this discrepancy is strongly reduced.     

\section{Conclusions}

In the framework of a two-particle RPA model applied to the description of
\bo and \nit formed of a correlated core of $^{12}$C   in its ground state
minus or plus a neutron-proton pair, we have performed a detailed analysis of
the T=0 effective nucleon-nucleon interaction. First  we have shown that for
the 0$^+$ T=1 states the results are as good as they were for $^{14}$C-$^{10}$C
(where the neutron-proton pair is replaced by a two-neutron pair) when the
Gogny interactions, D1 and D1S,  or their zero range equivalent are used. 
The same
 two-nucleon RPA model has been  applied to 1$^+$ and 3$^+$, T=0 states  and
 an attempt to analyse the contains of an effective interaction has been
 made.     We have shown that an effective
interaction constructed as a bare interaction, from Minnesota or of zero
range, fitted to the deuteron binding energy, complemented by medium effects
due to phonon exchanges between the neutron-proton pair gives a very good
representation of the \nit levels but fails to reproduce the 3$^+$ ground
state of \bo. By comparing with shell model calculations we are able to
suggest that this is due to the presence of a three-body component in the 
interaction which
 are not included in our derivation. This additionnal component 
 will not spoil our good results 
 in \nit but will improve those in \bo. At last our
study suggests that the T=0 Gogny interaction which yields good agreement
with measurements in both nuclei, includes empirically both the effect of
phonons exchange in the effective 
interaction and  the effect of a   
two-body component coming from the presence of
a three-body interaction and included  in the density dependent term. 

I am very grateful to E. Viggezzi and F. Barranco for pointing  to me a crucial
problem in the early stage of this work. Many thanks are also due to N.
Auerbach and B. Barrett for their interest in this work and for a careful
reading of the manuscript.

\appendix
\section{Matrix elements of the interaction induced by phonon exchange.}

The matrix elements of the induced interaction of Figure I are calculated
between two neutron-proton pairs coupled to spin J,M and isospin T=0. 

Each vertex in the diagrams is replaced by a phenomenological expression.
Let's call L,M$_L$ the angular momentum of the phonon, the transition density
from a zero-phonon state to the one-phonon state is written as:
\be
<1ph|V|0ph>=\frac{1}{\hat{L}}\beta_L R_0 \frac{dU(r)}{dr} Y_L^{M_L*}(\omega)
\ee
where U(r) is the average one-body potential   assumed to be a
Saxon-potential such as:
\be
\frac{dU(r)}{dr}=-U_0 \frac{df(r)}{dr}=\frac{U_0}{a} g(r)
\ee

where the function f(r) and the values of the parameters are given in 
section II. 
The collective amplitudes, $\beta_L$, can be fitted on the
experimental values of the B(EL) or on proton inelastic scattering cross
sections.  

The two-nucleon wave function is constructed by coupling the two-nucleon
states to a total spin (JM) as:

\be
|j_1 j_2,JM>=\sum_{m_1m_2}<j_1j_2m_1m_2|JM> |l_1j_1m_1>|l_2j_2m_2>
\ee
where the nucleons 1 and 2 are either in occupied states, $\alpha_1$
$\alpha_2$, or unoccupied states, $a_1$ $a_2$ with energies
$\epsilon_{\alpha_1}$, $\epsilon_{\alpha_2}$ or $\epsilon_{a_1}$
$\epsilon_{a_2}$ respectively. The total wave function has to
be antisymmetric   so that for T=0   states the spin-space wave function has
to be symmetric. Therefore the general expression for the
antisymmetrised matrix element of the induced interaction, $V_{ind}$, is
obtained as:
\be
<12|V_{ind}|34>=\sum_L\{ V_{dL}D_{dL}+V_{eL} D_{eL}\}
\ee

with:

\ba
V_{dL}&=&\frac{1}{2}[1+(-1)^{l_2+l_4+L}]
(-1)^{j_1+j_3+J}\frac{\beta_L^2R_0^2U_0^2}{4\pi a^2}
\frac{1}{\sqrt{(1+\delta_{12})(1+\delta_{34})}}\hat{\jmath_1}\hat{\jmath_2}
\hat{\jmath_3}
\hat{\jmath_4} \nonumber \\
  & & \left(\begin{array}{ccc}
  j_1&j_3&L\\
  0.5&-0.5&0
  \end{array}
  \right)
  \left(\begin{array}{ccc}
  j_2&j_4&L \\
  0.5&-0.5&0
  \end{array}
  \right)
  \left\{\begin{array}{ccc}
  j_1&j_2&J\\
  j_4&j_3&L
  \end{array}
  \right\} 
  <1|g|3><2|g|4> \\
V_{eL}&=&\frac{1}{2}[1+(-1)^{l_2+l_3+L}]
(-1)^{j_1-j_3}\frac{\beta_L^2R_0^2U_0^2}{4\pi a^2}
\frac{1}{\sqrt{(1+\delta_{12})(1+\delta_{34})}}\hat{\jmath_1}\hat{\jmath_2}
\hat{\jmath_3}
\hat{\jmath_4} \nonumber\\
  & & \left(\begin{array}{ccc}
  j_2&j_3&L\\
  0.5&-0.5&0
  \end{array}
  \right)
  \left(\begin{array}{ccc}
  j_1&j_4&L \\
  0.5&-0.5&0
  \end{array}
  \right)
  \left\{\begin{array}{ccc}
  j_2&j_1&J\\
  j_4&j_3&L
  \end{array}
  \right\} 
  <2|g|3><1|g|4> \\
<m|g|n>&=&\int_0^\infty g(r) \phi_{l_mj_m}^*(r)\phi_{l_nj_n}(r)r^2dr
\ea
where the $\phi$'s are the one-nucleon radial wave functions.

While the quantities $V_{dL}$ and $V_{eL}$ have the same expressions
whatever are the initial and final two-nucleon states, occupied or
unoccupied, the expressions of $D_{dL}$ and $D_{eL}$ which come from energy
denominators have to be derived in three different cases corresponding to
the diagrams of Fig.I:

If (12)=(a$_1$ a$_2$), (34)=(a$_3$ a$_4$)  (diagrams a) and b))
\ba
D_{dL}&=&[\Omega-(\epsilon_{a_2}+\epsilon_{a_3}+\omega_L)]^{-1}+
[\Omega-(\epsilon_{a_1}+\epsilon_{a_4}+\omega_L)]^{-1}\\
D_{eL}&=&[\Omega-(\epsilon_{a_1}+\epsilon_{a_3}+\omega_L)]^{-1}+
[\Omega-(\epsilon_{a_2}+\epsilon_{a_4}+\omega_L)]^{-1}
\ea

If (12)=(a$_1$ a$_2$), (34)=($\alpha_1$ $\alpha_2$) or the inverse (diagrams
c) and d)):
\ba
D_{dL}&=&-[\epsilon_{a_2}-\epsilon_{\alpha_2}+\omega_L]^{-1} 
-[\epsilon_{a_1}-\epsilon_{\alpha_1}+\omega_L]^{-1} \nonumber \\
D_{eL}&=&-[\epsilon_{a_1}-\epsilon_{\alpha_2}+\omega_L]^{-1} 
-[\epsilon_{a_2}-\epsilon_{\alpha_1}+\omega_L]^{-1}
\ea

It is straightforward to show that:
\be
<a_1 a_2|V_{ind}|\alpha_1 \alpha_2>=<\alpha_1 \alpha_2|V_{ind}|a_1 a_2>
\ee

If (12)=($\alpha_1$ $\alpha_2$), (34)=($\alpha_3$ $\alpha_4$) (diagrams e)
and f)):
\ba
D_{dL}&=&-[\Omega-\epsilon_{\alpha_2}-\epsilon_{\alpha_3}+\omega_L]^{-1}
-[\Omega-\epsilon_{\alpha_1}-\epsilon_{\alpha_4}+\omega_L]^{-1} \nonumber \\
D_{eL}&=&-[\Omega-\epsilon_{\alpha_1}-\epsilon_{\alpha_3}+\omega_L]^{-1}
-[\Omega-\epsilon_{\alpha_2}-\epsilon_{\alpha_4}+\omega_L]^{-1}
\ea

In these equations $\Omega$ is the eigenvalue of the RPA eqs.(5-6) and
$\omega_L$ is the energy of the phonon L. When we add these phonon exchange   
contributions to the bare matrix elements, the RPA equations become
non-linear and will be solved by iteration.

Note that the approximation of eq.(18) implies that one can defined an
equivalent two-body term which has to be added to the bare neutron-proton
interaction which is the sum of separable terms of  the following form:

\be
\delta V=\frac{d\rho(r)}{dr}\frac{d\rho(r')}{dr'}Y_L^M(\omega)Y_L^M(\omega')
\ee

\end{document}